\documentclass[a4paper]{aa}
\usepackage{graphicx}

\newcommand{\be}{\begin{equation}}
\newcommand{\ee}{\end{equation}}
\newcommand{\XRS}{1E~1024.0--5732}
\newcommand{\GRS}{3EG~J1027--5817}
\newcommand{\WRXXIa}{\hbox{WR~21a}}

\newcommand{\Einstein}{the {\it Einstein} Observatory}
\newcommand{\ASCA}{{\it ASCA}}
\newcommand{\RXTE}{{\it RXTE}}

\newcommand{\mC}[3]{\multicolumn{#1}{#2}{#3}}

\newcommand{\eg}{{\it e.g.}}

\newcommand{\ads}[2]{#2}
\newcommand{\Lx}{$L_{\rm X}(1-2~{\rm keV})$}
\newcommand{\Lu}{$10^{32}{\rm d}_{\rm kpc}^2 {\rm erg}{\rm s}^{-1}$}

\newcommand{\persec}{{\rm s}^{-1}}

\begin{document}

\title{Multiwavelength studies of \WRXXIa\ and its surroundings}

\author{Paula Benaglia\inst{1,2,}\thanks{Member of
           Carrera del Investigador, CONICET},
        Gustavo E. Romero\inst{1,2,*},
        B\"arbel Koribalski\inst{3},
        A. M. T. Pollock\inst{4}}

\institute{Instituto Argentino de Radioastronom\'{\i}a,
   C.C.5, (1894) Villa Elisa, Buenos Aires, Argentina \and
   Facultad de Cs. Astron\'omicas y Geof\'{\i}sicas, UNLP, Paseo del Bosque
   S/N, (1900) La Plata, Argentina \and
   Australia Telescope National Facility, CSIRO, PO Box 76, Epping,
     NSW 1710, Australia \and
   XMM-Newton Science Operations Centre, European Space Astronomy Centre,
     Apartado 50727, 28080 Madrid, Spain}

\offprints{P. Benaglia \\ \email{pbenaglia@fcaglp.unlp.edu.ar}}

\date{\today}

\titlerunning{\WRXXIa\ and its surroundings}
\authorrunning{Paula Benaglia et al.}

\abstract{ We present results of high-resolution radio continuum
observations towards the binary star \WRXXIa\ (Wack 2134) obtained
with the Australia Telescope Compact Array (ATCA\thanks{The
Australia Telescope is funded by the Commonwealth of Australia for
operation as a National Facility managed by CSIRO.}) at 4.8 and
8.64 GHz. We detected the system at 4.8 GHz (6 cm) with a flux
density of $0.25\pm0.06$ mJy and set an upper limit of 0.3 mJy at
8.64 GHz (3 cm). The derived spectral index of $\alpha < 0.3$
($S_{\nu} \propto \nu^\alpha$) suggests the presence of
non-thermal emission, probably originating in a colliding-wind
region. A second, unrelated radio source was detected $\sim10"$
north of \WRXXIa\ at (RA, Dec)$_{\rm J2000}=(10^{\rm h}25^{\rm
m}56^{\rm s}.49, -57^\circ48'34.4")$, with flux densities of 0.36
and 0.55 mJy at 4.8 and 8.64 GHz, respectively, resulting in
$\alpha = 0.72$. H\,{\sc i} observations in the area are dominated
by absorption against the prominent H\,{\sc ii} region RCW~49.
Analysis of a complete set of archived X-ray observations of
\WRXXIa\ confirms its strong variability but throws into doubt
previous suggestions by Reig (1999) of a period of years for the
system. Finally, we comment on the association with the
nearby EGRET source \GRS. }

\maketitle

\section{Introduction}

Massive O and Wolf-Rayet (WR) stars lose large amounts of mass through dense,
energetic winds. These winds can interact with the surrounding interstellar
medium (ISM) to create cavities in the H\,{\sc i} distribution (e.g. Benaglia
\& Cappa 1999) or produce strong shock fronts that, in principle, can
accelerate charged particles up to relativistic energies (e.g. V\"olk \&
Forman 1982). In fact, non-thermal radio emission has been detected in a
number of early-type stars, indicating the presence of populations of locally
accelerated electrons (e.g. Chapman et al. 1999, Benaglia et al. 2001b,
Benaglia \& Koribalski 2004).

Particle acceleration mediated by strong shocks could be taking place in
a number of different regions of a massive stellar system such as the outer
boundary of the interaction between the wind and the ISM (e.g. Cass\'e \&
Paul 1980), the base of the wind, where line-driven instabilities are thought
to drive strong shocks (\eg~White 1985), or, in massive binary systems, the
colliding-wind region (CWR, \eg~Eichler \& Usov 1993). In addition to the
expected non-thermal radio emission from relativistic electrons,
high-energy radiation could be produced through, for example,
inverse-Compton scattering of UV stellar photons or hadronic interactions
from shock-accelerated ions and ambient material (\eg~Pollock 1987, White
\& Chen 1992, Romero et al. 1999, Benaglia et al. 2001a, M\"ucke \& Pohl 2002,
Benaglia \& Romero 2003).

Only a small number of non-thermal radio WR binaries have been
identified so far (see for instance Dougherty et al. 2003) and,
even then, the connection between the radio and X-ray properties
of these systems is far from clear. The archetype system WR~140 is
bright in both regimes with a well-defined radio light curve that
shares the 7.94-year period of the optical radial-velocity orbit.
The X-ray light curve is not so well determined but shows strong
variability consistent with the same period. WR~140, however, is
the exception. The radio brightest star, WR~146, is a relatively
faint X-ray source whereas WR~147 seems to be intermediate in both
regimes.
Sharing some characteristics with these systems, the star \WRXXIa\
[($\alpha, \delta)_{\rm J2000} = 10^{\rm h}25^{\rm m}56^{\rm
s}.49, -57^\circ 48'44.4"$, $(l,\;b)=284^\circ.52,-0^\circ.24$],
is particularly interesting for various reasons. In the past, it
coincided within the fairly large positional uncertainties with
the COS~B source 2CG~284-00 (Goldwurm et al. 1987). More recently,
the unidentified EGRET gamma-ray source \GRS\ was detected
southwest of \WRXXIa\ (Hartman et al. 1999), being the star 30
arcmins apart from the center of the gamma-ray source. The 95\%
probability radius of the EGRET source is about 0.3$^\circ$.

Caraveo et al. (1989) established that \WRXXIa\ is associated with
the X-ray source \XRS, and reported X-ray pulses. Its X-ray
emission could be explained if the star is part of a binary system
with either a compact companion forming a high-mass X-ray binary
(Caraveo et al. 1989) or another massive, early-type star, with a
CWR (Reig 1999). According to its stellar spectrum, it was early
classified as an O5 star (Caraveo et al. 1989), and as a likely
binary (WN5-6+O3f), at a maximum distance of 3 kpc (Reig 1999).

With the current investigation we aim at determining the radio properties
of the star and its surroundings. Specifically, using interferometric
observations at 3 and 6 cm, we looked for non-thermal radiation that might
be interpreted as evidence for sites where electrons are being accelerated
in colliding winds or terminal shocks. By means of low-resolution 21cm-line
observations, we have also studied the distribution of neutral hydrogen
around the star. The detection of non-thermal emission coincident with the
stellar position would help in the identification of the system components.
Analysis of all available X-ray data of \WRXXIa\ reveals a variability
history that sheds fresh light on the nature of the binary. Finally, we
discuss the possibility of a physical link between the stellar system and the
nearby EGRET source, which can help to unveil the nature of the latter.

The contents of the paper are as follows: Section~2 reviews the
main sources detected in this region of the sky that are relevant
for the present study. Section~3 describes the observations
carried out and the data reduction; Section~4 outlines the new
observational results, whereas Section~5 presents the
corresponding analysis. In Section~6 we discuss the X-ray data.
Section~7 contains a comment on the possibility of a
physical association between the star and the EGRET source \GRS.
Section~8 closes with the summary.

\section{\WRXXIa\ and its surroundings}

The region of the Galactic plane towards $(l, b) = (284^{\circ},
0^{\circ})$ has been widely observed, from radio to gamma rays.
The optical images of Rodgers et al. (1960) revealed various
H\,{\sc ii} complexes, of which RCW~49 is the most prominent. The
COS~B satellite discovered the unidentified $\gamma$-ray source
2CG~284-00 (Bignami \& Hermsen 1983, Caraveo 1983). The zone was
then investigated by means of Einstein X-ray measurements in which
were detected emission from RCW~49 and a point source named \XRS\
(Goldwurm et al. 1987). The point source was tentatively linked to
\WRXXIa\ (Hertz \& Grindlay 1984), an emission-line star with $m_V
= 12.8$ (Wackerling 1969), also called Th35$-$42 and Wack~2134.

Caraveo et al. (1989) gathered enough evidence definitely to
associate the X-ray source with the star. They took optical
stellar spectra with the 3.6-m telescope at La Silla, classified
the star as O5, set an upper limit for the stellar distance of 3
kpc, and studied the Einstein X-ray emission, reporting an X-ray
periodicity of $\sim 60$ ms. They explained the X-ray behaviour as
a binary with a compact companion, likely an accreting neutron
star (NS). Dieters et al. (1990) looked for optical pulsations
with the Tasmania 1-m telescope. They set an upper limit around
19.7 mag for any optical pulsation.

Mereghetti et al. (1994) observed the region with the ROSAT PSPC
instrument, finding no pulsations, and obtained a further spectrum
with the CTIO 1.6m telescope, revising the spectral classification
to WN6. Based on the small equivalent widths of the emission lines,
they suggested that the object could be a binary with an O star,
rather than a compact companion, in which case the X-rays could come
from colliding winds, in common with other Wolf-Rayet binary systems.

In 1999, Reig presented RXTE data and a new spectrum taken with
the 1.9m telescope at SAAO. He stated that the lack of pulsations
and the relatively soft and low X-ray emission seem to exclude the
presence of a NS as responsible for the observed X-rays. Claiming
that the spectrum shows features of both WR and O stars, he
suggested that the system is formed by a WN6 and a possibly
supergiant O3 companion, favoring the hypothesis of the
colliding-wind binary (CWB). Roberts et al. (2001) observed the
region towards WR 21a with ASCA, and interpreted the hard X-ray
emission detected as produced by shocks from colliding winds in
the stellar system. Very recently, Niemela et al.'s (2005) optical
radial measurements have finally confirmed that the system is
formed by two massive stars.

The distance $d$ to WR 21a is not well established. We shall adopt
$d = 3$ kpc throughout this paper, a value that seems to be
consistent, as we will see, with all current observations.

Table~1 lists the adopted parameters that make up the Wack
2134/\WRXXIa\ binary system. Very few of the properties have so
far been measured, so we have to assume parameters from similar
stars or theoretical predictions.
The stellar luminosity, effective temperature, and stellar mass of
the WR component were estimated as an average of the same
variables given by Hamann et al. (1995) for WN6 stars. The WR mean
molecular weight of the ions $\mu$ was assumed the same as in
Leitherer et al. (1997) for a WN6 star. The WR wind terminal
velocity is taken as a mean value between data listed by Hamman et
al. (1995) and by Prinja et al. (1990) for WN6 stars. The WR
predicted mass loss rate ($\dot{M}$) is taken as the lowest value
tabulated for WN6 stars by Nugis \& Lamers (2000), in their
compilation of observable $\dot{M}$, and it can be considered a
lower limit. The O3 (I) mass was taken as the spectroscopic mass
from the tables of Vacca et al. (1996), and its terminal velocity
from the averaged values listed by Prinja et al. (1990). We
assumed $\mu= 1.5$ for the O star because of its evolved stage. A
predicted mass loss rate was estimated using the recipe derived by
Vink et al. (2000) (astro.ic.ac.uk/$\sim$jvink/). The expected
(WR+Of) mass-loss rate would imply a 6-cm flux density of 0.24 mJy
at 3 kpc, if thermal emission from ionized winds in both stars is
assumed.

Close to \WRXXIa\ there are two interesting sources: the H\,{\sc ii} region
RCW~49, and the gamma-ray source \GRS. They are discussed below.

\begin{table}
\caption[]{Adopted stellar parameters for the binary system \WRXXIa}
\begin{center}
\begin{tabular}{l r r l}
\hline Variable & WR& OB & unit \cr
\hline Spectral Class. & WN6$^{\rm (a)}$   & O3 (I)f$^{\rm (b)}$
&   \cr
log ($L/{\rm L}_\odot$) & 5.2$^{\rm (c)}$   & 6.274$^{\rm (d)}$
&   \cr
$T_{\rm eff}$           & 45000$^{\rm (c)}$ & 50680$^{\rm (d)}$
& K \cr $M_*$           & 12$^{\rm (c)}$    & 65$^{\rm (d)}$
& M$_\odot$\cr
$\mu$                   & 4$^{\rm (c)}$ & 1.5$^{\rm (c)}$
& \cr
$v_\infty$              & 2000$^{\rm (c)}$  &
3150$^{\rm (c)}$ & km s$^{-1}$\cr
$\dot{M}$ expected & $3.0 \times 10^{-5{\rm (c)}}$
& $1.4 \times 10^{-5{\rm (c)}}$ & M$_\odot$ yr$^{-1}$ \cr
distance $d$            & 3$^{\rm (c)}$  & 3$^{\rm (c)}$   & kpc \cr
\hline
\multicolumn{4}{l} {(a) van der Hucht 2001; (b) Reig 1999; (c) see
text;} \cr \multicolumn{4}{l} {(d) Vacca et al. 1996} \cr
\end{tabular}
\end{center}
\label{table1}
\end{table}

\subsection{The  H\,{\sc ii} region RCW~49}

RCW~49 is a southern H\,{\sc ii} region, located at $(l,b) =
(284.3^{\circ}, -0.3^{\circ})$, and extended over an area of
$\sim90' \times 70'$. Values for its distance range between 2.3
and 7.9 kpc (Manchester et al. 1970, Moffat et al. 1991, Brand \&
Blitz 1993, etc). H\,{\sc i} spectra towards RCW~49 taken by Goss
et al. (1972) show prominent H\,{\sc i} absorption from about --20
to +5 km\,s$^{-1}$ (see also Figs.~2 and 3). Whiteoak \& Uchida
(1997) have imaged RCW~49 at radio continuum with MOST at 0.843
GHz, and the central region using ATCA at 1.38 and 2.38 GHz,
attaining an angular resolution of 7 arcsec. They found two
shells, and ascribe the formation of the northern one to the
Westerlund 2 cluster containing the binary star WR~20a, and the
southern one to the star WR~20b (see their Fig. 2d). Recent
infrared images of RCW~49 obtained with the Spitzer Space
Telescope (Churchwell et al. 2004) show the intricate filamentary
structure of the nebula in the inner 5 arcmin shaped by stellar
winds and radiation.

\subsection{The gamma-ray source \GRS}

After the analysis of the EGRET data, Hartman et al. (1999) were
the first to point at the proximity between the gamma-ray source
\GRS\ and the X-ray source \XRS, associated with \WRXXIa\ (see
Figure 3). The averaged measured flux at $E>100$ MeV is
65.9$\pm0.70 \times 10^{-8 }$ photons cm$^{-2}$ s$^{-1}$. The
source is constant within errors on timescales of months
(variability index $I$ = 1.6, see Torres et al. 2001) and with a
photon spectral index $\Gamma = 1.94\pm0.09$ ($dN/dE\propto
E^{-\Gamma}$). The more recent variability analysis by Nolan et
al. (2003), who calculated a likelihood function for the flux of
each source in each observation, also suggests that this source is
not variable on short, monthly timescales.

\section{Observations and data reduction}

In order to search for non-thermal radio emission in the direction
of \WRXXIa\ we have carried out interferometric radio continuum
observations at high angular resolution ($1'' - 2''$). These were
complemented with single-dish H\,{\sc i}-line observations to map
the distribution of neutral material in the neighbourhood and
study its kinematic behaviour.

\subsection{Radio continuum observations}

Radio continuum data were obtained in September 2001 with the
Australia Telescope Compact Array in the 6D array, observing
simultaneously at 3 and 6 cm, or 8.64 and 4.8 GHz, respectively.
The total bandwidth used was 128 MHz. The primary calibrator was
PKS~1934-638, with flux densities of 5.83 and 2.84 Jy, at 4.8 and
8.64 GHz. \WRXXIa\ was tracked 12 h --full synthesis-- to gain
maximum $uv$ coverage, interleaving with observations of the phase
calibrator 1039--47. The theoretical r.m.s. noise after $\sim$ 9 h
on source is 0.03 mJy at both frequencies, taking all baselines into
account.

The data were reduced and analyzed with {\sc miriad} routines. The
images built using ``robust'' weighting showed the best signal to
noise ratio, and minimized sidelobes. The diffuse emission from
extended sources was removed by taking out the visibilities
corresponding to the shortest baselines. The resulting beams were
$0.83" \times 0.70"$  at 3 cm, and $1.73" \times 1.49"$ at 6 cm.
The r.m.s. noise of the final maps is 0.1 mJy beam$^{-1}$ at 3 cm
and 0.06 mJy beam$^{-1}$ at 6 cm.  The difference between the
theoretical and the observed r.m.s. noise is mainly due to flagging
of bad data -- specially at 3 cm -- as well as short baselines
contributing with confusing emission from extended sources both in
and outside the main beam.

The observations were set to optimize the detection of point-like features.
Maps at two frequencies would allow the determination of spectral indices.

\begin{figure*}
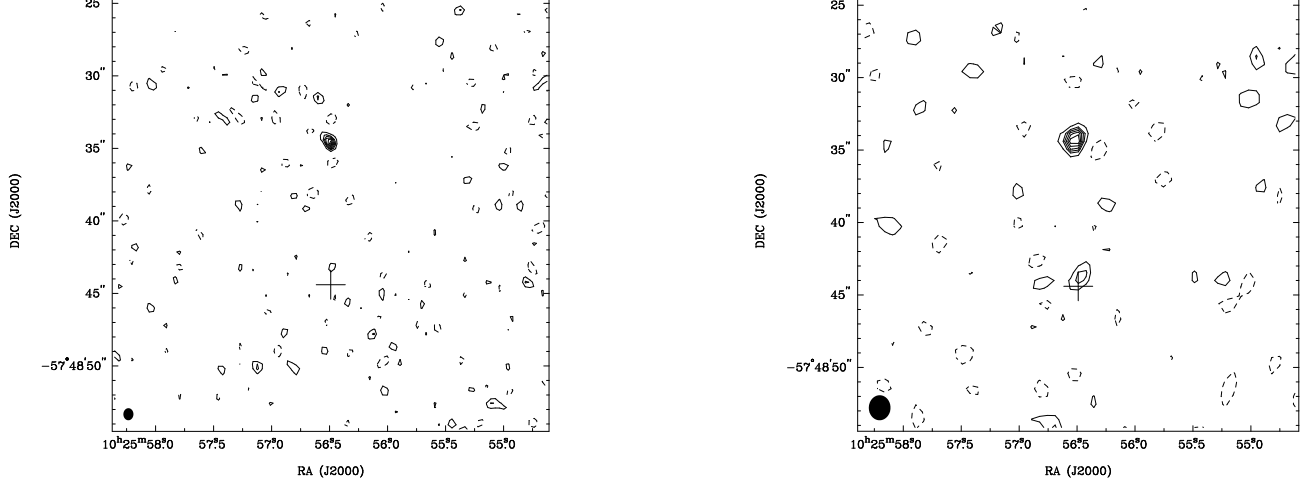
 
\centering
 \begin{minipage}{0.55\textwidth}
  \includegraphics[width=65mm,angle=-90]{2617fig1.ps}
 \end{minipage}%
 \begin{minipage}{0.50\textwidth}
  \includegraphics[width=65mm,angle=-90]{2617fig2.ps}
 \end{minipage}%
\caption{{\it Left panel}: ATCA-3cm radio continuum image towards
    \WRXXIa. The optical position of the star is marked with a cross.
    The contour levels are --0.20, 0.20 ($2\sigma$), 0.30, 0.35, 0.40,
    0.45, 0.50, and 0.55 mJy\,beam$^{-1}$. The synthesized beam ($0.83"
    \times 0.70"$) is displayed at the bottom left corner.
    {\it Right panel}: ATCA-6cm radio continuum image towards \WRXXIa.
    The contour levels are --0.12, 0.12 ($2\sigma$), 0.18, 0.24, 0.30,
    and and 0.33 mJy\,beam$^{-1}$. The synthesized beam ($1.73" \times
    1.49"$) is displayed at the bottom left corner.}
\end{figure*}

\subsection{ H\,{\sc i}-line observations}

The 21cm-line data were obtained with a 30m-single dish
radiotelescope at the Instituto Argentino de Radioastronom\'{\i}a
(IAR, Villa Elisa, Argentina) in June 2000. The observations were
done in total power mode, covering a total field of $5^{\circ}
\times 5^{\circ}$, in Galactic coordinates with a cell size of
$15' \times 15'$. The HPBW at 1420 MHz is 30'. The receiver's
system temperature was $\approx$ 35 K. The velocity coverage was
(--450, +450) km\,s$^{-1}$; the 1008 channel autocorrelator
allows a maximum velocity resolution of 1.05 km\,s$^{-1}$. The r.m.s.
noise of the brightness-temperature ($T_{\rm B}$) of a single spectral
point is $\sim$ 0.1 K. The $T_{\rm B}$ scale was calibrated with the
standard region S9 (Morras \& Cappa 1995). A series of
($l$, $b$)-$T_{\rm B}$ maps were built every 1.05 km\,s$^{-1}$ to
proceed with the analytical stage.

\section{Results}
\subsection{ATCA radio continuum data}

The ATCA images at 3 and 6 cm are shown in Figure 1. A point source
positionally coincident with \WRXXIa\ is visible at 6 cm, at $S/N$ $>$ 4.
A flux density $S_{\rm 6cm} = 0.25$ mJy was derived using {\sc imfit}
after a gaussian fit (see Table 2). The r.m.s. noise in the 3 cm image
is 0.1 mJy beam$^{-1}$. No radio source is detected at the stellar
position above 3 r.m.s.. This non-detection imposes an upper limit for
the spectral index of $\alpha < 0.3$ ($S \propto \nu^\alpha$). The
deviation from purely thermal emission, characterized by $\alpha = 0.6
- 0.8$, indicates the presence of a non-thermal contribution.

\begin{table*}
\caption[]{Radio continuum results}
\begin{center}
\begin{tabular}{lrccr}
\hline
Object & $S_{\rm 3cm}$ &  $S_{\rm 6cm}$ & RA,Dec (J2000) & $\alpha$\\
       & (mJy)            & (mJy)        & (hms, dms) & \\
\hline
\WRXXIa &  $< 0.3$ & $0.25\pm0.06$ & 10:25:56.49 $-$57:48:43.3 & $< 0.3$ \\
&&&&\\
S2 & $0.55\pm0.10$ & $0.36\pm0.06$ & 10:25:56.49 $-$57:48:34.4& 0.72 \\
\hline
\end{tabular}
\end{center}
\label{table2}
\end{table*}

In a similar way as in Benaglia \& Koribalski (2004), we derive a
mass loss rate of \WRXXIa, from the flux density at 6 cm. In a
first approximation, the mean number of electrons per ion
($\gamma$), and the rms ionic charge ($Z$) were taken equal to
unity. The gas temperature is computed from the stellar effective
temperature (see Table 1) as $0.4\, T_{\rm eff}$. The Gaunt factor
results in 5.6. At the adopted distance of 3 kpc,  this would
imply a radio-derived mass loss rate for the WR star of $\dot{M} =
f\,\times \,4.8\, \times \,10^{-5}$ M$_{\odot}$ yr$^{-1}$, where
$f$ is the fraction of thermal to total radio emission. From
the relation between the flux density values measured with ATCA at
3 and 6 cm we know that at 6 cm there is a non-thermal
contribution to the emission. Thus, in deriving a mass loss rate
from the flux density at 6 cm, the result with $f = 1$ is an upper
limit.

A second source, called S2, is detected 10'' away from \WRXXIa, at
both frequencies. The measured fluxes are 0.55 mJy at 3 cm, and
0.36 mJy at 6 cm. The corresponding spectral index of $\alpha_{\rm
S2} = +0.72$ points to an ultracompact  H\,{\sc ii} region or the
thermal wind of another, as yet unidentified, early-type star on
the field of \WRXXIa. Table 2 lists the position and flux density
of the two sources detected.

\subsection{Large-scale H\,{\sc i}-line data}\label{HI}

Studies of neutral hydrogen were performed to look for evidences
of gas features such as shells, filaments, bubbles, etc., that
could be physically related to the object under the present study.
The H\,{\sc i} gas kinematics in the line-of-sight to a Galactic
source can be used to constrain its distance (\eg~Koribalski et
al. 1995) using the Galactic rotation curve (\eg~Fich et al. 1989)
and the Galactic velocity field (Brand \& Blitz 1993). Here we
searched for the signatures of an interstellar H\,{\sc i} bubble,
created by the action of the stellar winds of \WRXXIa. Because of
the large angular size of the IAR telescope beam (HPBW = 30
arcmin; see Figs.~2 and 3) and the proximity of \WRXXIa\ to the
extended H\,{\sc ii} region RCW~49, which has a total radio
continuum flux of $\sim$210 Jy at 843 MHz (Whiteoak \& Uchida
1997), H\,{\sc i} spectra in this direction are completely
dominated by H\,{\sc i} absorption against RCW~49 (Goss et al.
1972; McClure-Griffiths et al. 2001). We also investigated the
high-resolution (130 arcsec) H\,{\sc i} data cubes from the
Southern Galactic Plane Survey (SGPS; McClure-Griffiths et al.
2001) in the region of RCW~49 and WR~21a. Unfortunately, the
region around RCW~49, including the H\,{\sc i} line emission at
the position of WR~21a, suffers from artifacts caused by the
strong radio continuum emission from RCW49 and potential H\,{\sc
i} structures associated with WR~21a cannot be distinguished.

Figure 2 displays the  H\,{\sc i} brightness temperature maps each
4 km s$^{-1}$ built from the IAR data.

If we were able to see an H\,{\sc i} bubble around WR~21a, its
size would give us some information about the energetics of the
stellar wind and its velocity would give us an estimate of its
kinematic distance using the Galactic rotation curve and velocity
field. These issues are explored in the following section.

\begin{figure*}[] 
\begin{center}
\resizebox{15cm}{!}{\includegraphics{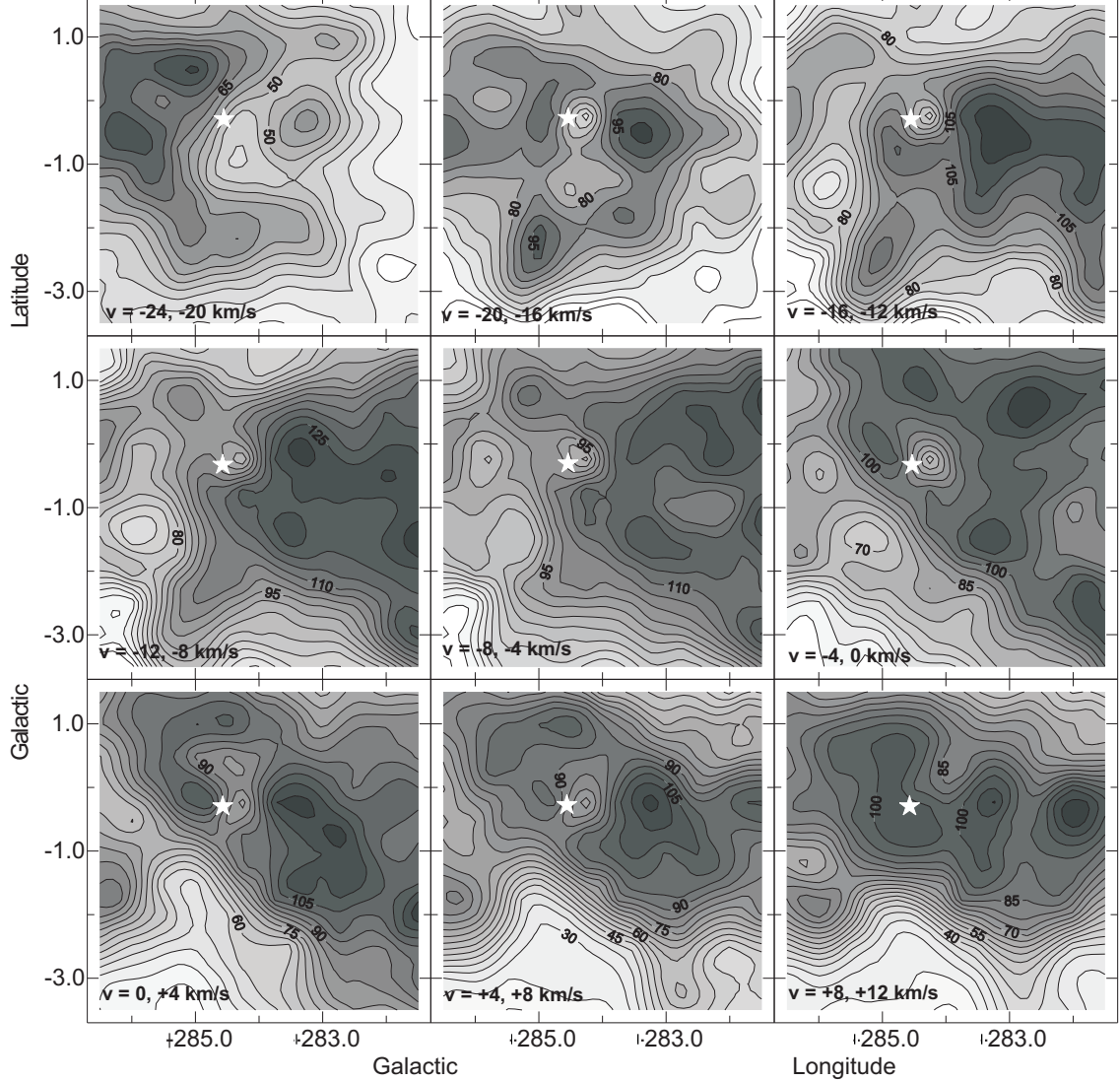}}
\caption{H\,{\sc i}-brightness temperature distribution of the Galactic
  emission as measured with the IAR telescope (HPBW = 30 arcmin) over a
  velocity range from --24 to +12 km\,s$^{-1}$, in steps of 4 km\,s$^{-1}$.
  The position of WR 21a is marked with a white star. The contour levels
  indicate H\,{\sc i} brightness temperatures in steps of 5 Kelvin.}
\label{figure2}
\end{center}
\end{figure*}

\begin{figure}
\begin{center}
\resizebox{8cm}{!}{\includegraphics{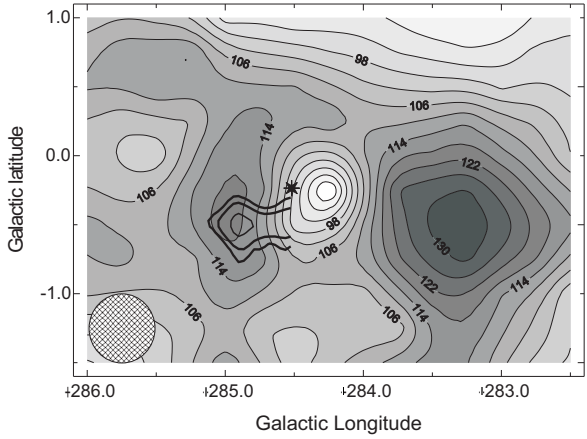}}
\caption{Neutral-hydrogen column density integrated over
velocities from
  --21 to --14 km\,s$^{-1}$. The contour levels indicate H\,{\sc i} brightness
  temperatures in steps of 4 Kelvin. The IAR telescope beam is displayed in
  the bottom left corner. The position of \WRXXIa\ is marked with a
  black
  star, and the black contours represent the 99, 95 and 50\% probability
  contours for the location of the gamma-ray source \GRS.}
\label{figure3}
\end{center}
\end{figure}

\section{Analysis of the line features}

The derivation of gas kinematic distances is difficult for this
particular region of the Galactic plane. The line of sight goes
tangential to the Carina arm at this Galactic longitude, and
velocity crowding becomes very important. Brand \& Blitz (1993)
showed that the measured velocities deviate strongly from Galactic
rotation: gas in the velocity range from --20 to --10 km\,s$^{-1}$
can be located at distances between 2 and 6 kpc for $l \sim
285^\circ$, where both \WRXXIa\ and RCW~49 are likely to be
located.

By means of CO observations, Grabelsky et al. (1987) studied
molecular gas associated with the Carina arm. They interpreted the
velocity-longitude behaviour of the gas in terms of material at
different heliocentric distances (see their Fig. 5), separating
local clouds from Carina arm gas. According to their results it is
possible to determine that toward $l \sim 284^\circ$ gas showing
velocities around $\sim-$15 km\,s$^{-1}$ belongs to the Carina
arm and is located at about 3 kpc, i.e., the distance of WR 21a.
In Fig. 2 it can be appreciated that the gas distribution changes
at about $v \sim -14$ to --12 km\,s$^{-1}$. We are going to focus
on gas with $\Delta v$ = --20 to --12 km\,s$^{-1}$ because,
according to the CO results, its kinematical distance is
compatible with that of the target star. The H\,{\sc i} column
density at the mentioned velocity interval is presented in Fig.~3.
From the plot of Grabelsky et al. (1987), CO gas with velocities
between --20 and --10 km\,s$^{-1}$ is located between 2 and $\sim
3.5$ kpc. This fact helps to constrain an approximate error in the
H\,{\sc i} distance of $\leq$ 1 kpc. At larger velocities,
Grabelsky et al. claimed that gas related to RCW~49 shows
velocities of --5 km\,s$^{-1}$, and placed it at 4 kpc. A distance
of $\sim$5 kpc can be derived, for Carina gas with velocities near
0 km\,s$^{-1}$. If the H\,{\sc i} follows the motions proposed by
Grabelsky et al., it is reasonable to suggest that gas with a
velocity of --14 km\,s$^{-1}$ lies at 3 kpc.

Figure 3 shows the presence of gas at a distance compatible with
that of the stellar system. Due to the strong
continuum source RCW 49, part of the HI gas is not emitting but
absorbing. At a distance of 3 kpc, the visible neutral gas
coincident with the position of the EGRET source would sum up
about 1500 $M_\odot$, which can be considered as a lower limit for
the masses of clouds at the distance we are interested on.

By means of Antenna~I at IAR (HPBW = 12.4$^\circ$) we also
measured the H125$\alpha$ (3326.9880 MHz) radio recombination line
(RRL) towards $(l,b) = (284.32^\circ, -0.34^\circ)$. Since RRLs
are typically produced by H\,{\sc ii} regions, the detected
emission (see Fig.~4) is likely to be mostly from RCW~49. We
measure a center velocity of $v = +0.6\pm1.4$ km\,s$^{-1}$,
similar to those found at other RRLs in RCW~49 (\eg~Caswell \&
Haynes 1987).

\begin{figure} 
\centering
\includegraphics[width=7cm]{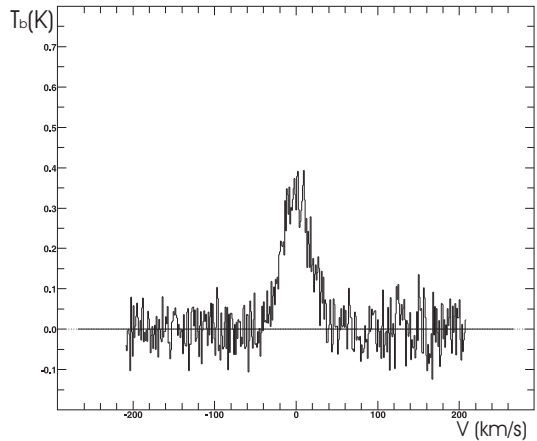}
\caption{3.3 GHz radio recombination line H125$\alpha$ towards RCW 49.
  HPBW = 12.4$^{\circ}$.}
\label{figure4} 
\end{figure}

The kinetic energy needed to form a typical bubble around WR~21a
can be computed as $E_{\rm k} = 0.5 M_{\rm sh} v_{\rm exp}^2 =
10^{49}$ erg if we assume a shell mass of $M_{\rm sh}$ = 10000
M$_\odot$ and an expansion velocity of $v_{\rm exp}$ = 10
km\,s$^{-1}$, which are typical parameters of neutral shells
detected around massive Of stars (Cappa \& Benaglia 1998, Benaglia
\& Cappa 1999).

The stellar wind luminosity can be expressed as $L_{\rm w} = 0.5
\dot{M} v_{\infty}^2$. Using the values of $\dot{M}$ and
$v_{\infty}$ given in Table~1, we find $L_{\rm w,WR} = 3.8 \times
10^{37}$ erg\,s$^{-1}$, and $L_{\rm w,O}  = 4.4 \times 10^{37}$
erg\,s$^{-1}$.

Finally, a wind mechanical energy $E_{\rm w} = L_{\rm w} \, t =
2.6 \times 10^{51}$ erg  per Myr is obtained if both stars are
considered, and $E_{\rm w} = 1.2 \times 10^{51}$ erg per Myr for
only WN6. It can be seen that the energy deposited by the wind is
much larger than the energy needed to create a typical bubble.

\section{The X-ray history of \WRXXIa}

Among the Wolf-Rayet stars, \WRXXIa\ is especially prominent in
X-rays: despite its modest optical magnitude, it is one of the
five or six brightest sources in both apparent and absolute terms.
Since the discovery of \XRS\ in 1979 with \Einstein\ by Goldwurm
et al. (1987), through its identification with a Wolf-Rayet star
by Mereghetti et al. (1994), \WRXXIa\ has been observed on ten
separate occasions in X-rays as shown in Table~\ref{X-ray:log}.
All of these data are available through the HEASARC. As we
mentioned before, Reig (1999) compared his 1997-\RXTE\ data with
earlier Einstein-IPC and ROSAT-PSPC measurements to show that the
X-ray luminosity had apparently been steadily rising by about a
factor of five in the eighteen years since its discovery and
argued that this showed \WRXXIa\ is a long-period CWB of the type
exemplified by WR 140 (Williams et al. 1990).

Though in general terms the latter is probably correct, the
details are more complicated. First, the X-ray model which Reig
used had too high an absorbing column density ($N_{\rm
H}=2.9\times10^{22}{\rm cm}^{-2}$) to be consistent with the the
soft X-ray data; and second, the faintness of the pair of
ROSAT-HRI measurements made in 1994 (Belloni \& Mereghetti 1994)
contradicts the apparent inexorable rise in luminosity since 1979.
Since the \RXTE\ observation, two archived sets of ASCA data have
also become available that are useful for bridging the soft X-ray
images and the hard X-ray collimator data. The first ASCA
observation took place two months after Reig's RXTE pointing
though \WRXXIa\ was not the main objective and thus appeared near
the edge of the GIS field-of-view, one of the two ASCA instruments
that provided imaging X-ray spectroscopy. As a consequence there
are no data from the SIS, the other instrument. On the other hand,
data are available from the full set of ASCA instruments from an
observation performed nearly a year later at the end of 1998.

We have obtained and analysed with XSPEC v11.2 all the archived
spectra and associated response matrices available from the
HEASARC. Despite the obvious changes in overall luminosity, we
could find no evidence of any changes in the shape of the spectrum
which, within the limited energy resolution available, seems to be
consistent with the relatively hot few keV thermal plasma observed
from the Wolf-Rayet binaries exemplified by WR~140 (Pollock et al.
2005), in contrast with the cooler temperatures more typical of
the intrinsic emission of single stars (see for example the
studies on the presumably single WN stars WR 1 [Ignace et al.
2003], WR 6 [Skinner et al. 2002b], and WR 110 [Skinner et al.
2002a]). The spectrum was modeled as an absorbed Bremsstrahlung
continuum with additional emission lines of Si, Mg and Ne. The
best-fit values of column density and temperature were $N_{\rm
H}=7.0\pm0.6\times10^{21}{\rm cm}^{-2}$, about a factor of 4 lower
than Reig's (1999) value, and $kT=3.3\pm0.2$ keV. This empirical
approach has the natural advantage of reproducing the range of
ionization species in the spectrum, notably the simultaneous
presence of the lines of SiXIII and SiXIV. The alternative fits
given by XSPEC's plasma models were slightly worse but gave
completely consistent best-fit values of column density and
temperature.

\begin{figure} 
\begin{center}
\resizebox{8cm}{!}{\includegraphics{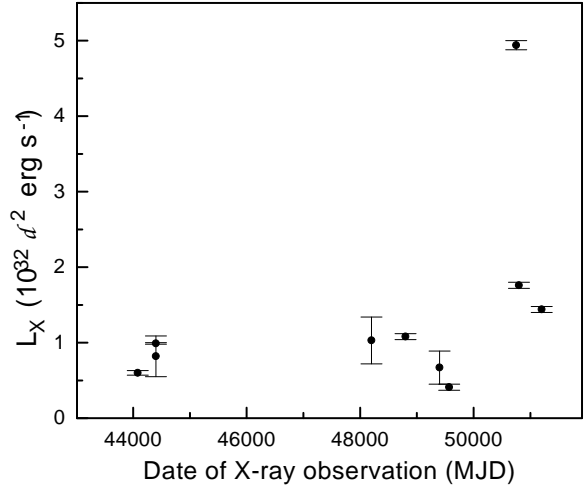}}
\caption{\label{Figure:X-ray:history}
   The history of \WRXXIa's 1-2~keV X-ray luminosity since its discovery in
   1979 with the Einstein Observatory, through measurements in the early
   1990s with ROSAT to the \RXTE\ and \ASCA\ measurements of 1997 and 1998.
   \WRXXIa\ is bright and the luminosity errors are usually smaller than
   the plotting symbols. The luminosities were calculated assuming the
   spectrum did not change shape and the range 1-2 keV was chosen to provide
   good overlap between instruments whose sensitivity ranges were different.
   The highest point is from \RXTE\ which, as explained in the text, probably
   included a significant contribution from nearby unresolved point-source
   and diffuse emission.
}
\end{center}
\label{[figure5}
\end{figure}

The luminosities reported in Table~\ref{X-ray:log} are for a joint
fit to all the available spectra with only the luminosities free
to vary between observations. The resulting lightcurve is shown in
Figure~\ref{Figure:X-ray:history}. The X-ray variability is
apparently irregular though the measurements are spaced at such
large intervals with respect to the newly-discovered period of
weeks (Niemela et al. 2005) that it will only be possible to tell
if it is related to the binary orbit once a precise orbit is
available. Some care is also required with the brightest point
that came from \RXTE, the only instrument here with no imaging
capabilities. Though Reig made a correction for the emission from
other nearby sources by adding an extra spectral component with
fixed parameters, this is quite uncertain, because of the
extensive diffuse emission and the strength of the point sources
enumerated by Belloni \& Mereghetti (1994), of which 1E1022.2-5730
is of similar spectral shape to \WRXXIa.

\begin{table*}
\caption{\label{X-ray:log} X-ray observations of \WRXXIa}
\begin{center}
\begin{tabular}{lllrl@{$\pm$}lc}
\hline Date       & Dataset ID               & Instrument      &
Time          & \mC{2}{c}{Count Rate} & \Lx  \\
           &                          &                 & ($s$)         & \mC{2}{c}($\persec$)  &(\Lu) \\
\hline  %
1979-07-13 & \ads{einstein}{3341}     & Einstein-IPC    &  1324
& 0.036  & 0.002                   &  0.60$\pm$0.03  \\
1980-07-08 & \ads{einstein}{7715}     & Einstein-IPC    &  7680
& 0.0592 & 0.0002                  &  0.99$\pm$0.01  \\
1980-07-10 & \ads{einstein}{7716}     & Einstein-HRI    &  7006
& 0.006  & 0.002                   &  0.82$\pm$0.27  \\
1990-10-01 & \ads{rosat}{RS932717N00} & RASS            &   180
& 0.0778 & 0.0233                  &  1.03$\pm$0.31  \\
1992-07-29 & \ads{rosat}{RP400329N00} & ROSAT-PSPC      &  8396
& 0.0817 & 0.0031                  &  1.08$\pm$0.04  \\
1994-01-14 & \ads{rosat}{RH201611N00} & ROSAT-HRI       &   696
& 0.0177 & 0.0059                  &  0.67$\pm$0.22  \\
1994-07-31 & \ads{rosat}{RH201611A01} & ROSAT-HRI       & 17906
& 0.0108 & 0.0012                  &  0.41$\pm$0.04  \\
1997-11-29 & \ads{rxte}{301120101}    & RXTE-PCA        & 12000
& 7.088  & 0.079                   &  4.94$\pm$0.06  \\
1998-01-21 & \ads{asca}{26014000}     & ASCA-GIS2       & 36858
& 0.0518 & 0.0012                  &  1.76$\pm$0.04  \\
1998-12-27 & \ads{asca}{46009000}     & ASCA-GIS2       & 21174
& 0.0817 & 0.0020                  &  1.44$\pm$0.04  \\
           &                          & ASCA-SIS0       & 21788     & 0.1208 & 0.0024                  &                 \\
\hline \\ \\
\end{tabular}
\end{center}
\end{table*}

\section{Connection with the EGRET source?}

The nature of the EGRET unidentified gamma-ray sources (Hartman et
al. 1999) has become one of the most intriguing questions in
astrophysics ({\it e.g.} Romero 2001). In the particular case of 3EG J1027-5817,
the nearby X-ray source associated with WR~21a is indicated in the Third 
EGRET catalog as a potential counterpart. 

The presence of a non-thermal contribution to the radio spectrum
of \WRXXIa\ implies the existence of a population of relativistic
particles in the source, which are probably accelerated at the CWR
by diffusive shock acceleration (\eg~Bell 1978a,b) and cool by
synchrotron emission in the local magnetic field. Electrons can
also cool in such an environment through inverse-Compton
interactions with stellar UV photons, producing non-thermal X-rays
and gamma-rays (\eg~Pollock 1987, Benaglia \& Romero 2003).
However, the facts that the X-ray emission from the system can be
correctly modeled as thermal Bremsstrahlung and that the source
has not been directly detected at gamma-rays suggest that the
magnetic energy density in the CWR should largely exceed the
photon energy density, implying significantly shorter cooling
timescales for the synchrotron mechanism.

The same mechanism that accelerates the electrons should
also operate on the ions. Synchrotron losses are not relevant for
protons in the environment of the CWR. The maximum energy they can
achieve will be determined by the photo-pion losses in the UV
stellar field and by the size constraint imposed by the limited
space available for the acceleration process. In the present case,
where the CWR is not resolved and the geometry of the system
remains unknown, we cannot calculate the high-energy cutoff for
the non-thermal proton distribution. A value between 10 and 100
GeV seems not unreasonable (see Benaglia \& Romero 2003).

It could be the case that some of these protons diffuse up to a
nearby cloud where they might be trapped in the magnetic field,
which is expected to be higher than the average value in the ISM
(Crutcher 1999). Then they will interact there with the local
material producing gamma-rays through $p+p\rightarrow p+p+\pi^0$
interactions and the subsequent $\pi^0\rightarrow \gamma+\gamma$
decays. The situation of a passive cloud irradiated by cosmic rays
from some nearby accelerator has been discussed in detail by Black
\& Fazio (1973) and by Aharonian \& Atoyan (1996). In the present
case our ignorance on several basic parameters prevents accurate
calculations, but there remains the possibility that a
part of the flux detected from 3EG J1027-5817 could be originated
in relativistic particles accelerated in the colliding wind region
of  WR 21a. Whether this is or not the case could be established
through future observations of the gamma-ray source by instruments
like AGILE and GLAST which could report the source position with
higher accuracy.

\section{Summary and Conclusions}

We have detected radio emission from \WRXXIa\ at 4.8 GHz. The
intensity of the source is $\sim$0.25 mJy. The non-detection at
8.64 GHz implies a spectral index of $\alpha<0.3$ ($S_{\nu}
\propto \nu^{\alpha}$), which significantly departs from a typical
Bremsstrahlung spectrum. Combined thermal/non-thermal spectra are
usually found in colliding-wind binaries. We suggest that this is
also the case here, since the latest spectral determinations show
that \WRXXIa\ is a system formed by WN6 and an early O companion
(Reig 1999, Niemela et al. 2005). An upper limit for the system
mass loss rate of $\dot{M} = 4.8 \times 10^{-5}$ M$_\odot$
yr$^{-1}$ is derived from the 4.8-GHz radio flux density.

We have reanalyzed all X-ray observations of \WRXXIa\ in order to determine
its time history in this waveband. Our results indicate, contrary to previous
thought, that the X-ray flux has not been monotonically increasing since 1979
though the coverage is far too sparse to constrain the variability timescale.
The X-ray spectrum is consistent with a few keV thermal plasma, with no obvious
non-thermal contribution.

Locally accelerated relativistic electrons in the CWR 
probably mainly cool by synchrotron emission at radio frequencies,
with small inverse-Compton losses, of which there was no X-ray
evidence. If protons are also accelerated at the colliding-wind
shocks, then they might diffuse through the ISM up to nearby
clouds, where they might interact with an enhanced  H\,{\sc i}
density to produce gamma-rays from $\pi^0$-decays.

Future observations with both X-ray and gamma-ray instruments like CHANDRA,
AGILE and GLAST can shed additional light on the nature of the high-energy
emission in this interesting region. Detailed knowledge, on the other hand,
of the orbital parameters of the system \WRXXIa\ will allow more sophisticated
models to be built of the radiative processes taking place in the
colliding-wind region.

\begin{acknowledgements}
We thank Virpi Niemela for discussions on this source. This research has been
supported by the Argentine agency ANPCyT through grant PICT 03-13291.
\end{acknowledgements}

{}


\begin{thebibliography}{}

\bibitem{}Aharonian, F.A., \& Atoyan, A.M. 1996, A\&A, 917, 928
\bibitem{}Bell, A.R. 1978a, MNRAS, 182, 147
\bibitem{}Bell, A.R. 1978b, MNRAS, 182, 443
\bibitem{}Belloni, T., \& Mereghetti, S. 1994, A\&A, 286, 935
\bibitem{}Benaglia, P., \& Cappa, C.E. 1999, A\&A, 346, 979
\bibitem{}Benaglia, P., Romero, G.E., Stevens, I.R., \& Torres, D.F. 2001a,
  A\&A, 366, 605
\bibitem{}Benaglia, P., Cappa, C.E., \& Koribalski, B.S. 2001b, A\&A, 372, 952
\bibitem{}Benaglia, P., \& Romero, G.E. 2003, A\&A, 399, 1121
\bibitem{}Benaglia, P., \& Koribalski, B. 2004, A\&A, 416, 171
\bibitem{}Bignami, G.F., \& Hermsen, W. 1983, ARA\&A, 21, 67
\bibitem{}Black, J.H., \& Fazio, G.G. 1973, ApJ, 185, L7
\bibitem{}Brand, J., \& Blitz, L. 1993, A\&A, 275, 67
\bibitem{}Cappa, C.E., \& Benaglia, P. 1998, AJ, 116, 1906
\bibitem{}Caraveo, P.A. 1983, Space Sci. Rev., 36, 207
\bibitem{}Caraveo, P.A., Bignami, G.F., \& Goldwurm, A. 1989, ApJ, 338, 338
\bibitem{}Cass\'e, M., \& Paul, J.A. 1980, ApJ, 237, 236
\bibitem{}Caswell, J.L., \& Haynes, R.F. 1987, A\&A, 171, 261
\bibitem{}Chapman, J.M., Leitherer, C., \& Koribalski, B.S., et al. 1999,
  ApJ, 518, 890
\bibitem{}Churchwell, E., Whitney, B.A., Babler, B.L., et al. 2004, ApJS 154, 322
\bibitem{}Crutcher, R.M. 1999, ApJ, 520, 706
\bibitem{}Dieters, S.W., Hill, K.M., \& Watson, R.D. 1990, IAU Inf. Bull., 3500
\bibitem{}Dougherty, S.M., Pittard, J.M., Kasian, L., et al. 2003, A\&A, 409 217
\bibitem{}Eichler, D., \& Usov, V. 1993, ApJ, 402, 271
\bibitem{}Fich, M., Blitz, L., \& Stark, A\&A, 1989, ApJ, 342, 272
\bibitem{}Grabelsky, D.A., Cohen, R.S., Bronfman, L., et al. 1987, ApJ, 315, 122
\bibitem{}Goldwurm, A., Caraveo, P.A., \& Bignami, G.F. 1987, ApJ, 322, 349
\bibitem{}Goss, W.M., Radhakrishnan, V., Brooks, J.W., \& Murray, J.D. 1972,
  ApJS, 24, 123
\bibitem{}Hamann, W.-R., Koesterke, L., \& Wessolowski, U. 1995, A\&A, 299, 151
\bibitem{}Hartman, R.C., Bertsch, D.L., Bloom, S.D., et al. 1999, ApJS, 123, 79
\bibitem{}Hertz, P., \& Grindlay, J.E. 1984, ApJ, 278, 137
\bibitem{}Ignace, R., Oskinova, L. M., \& Brown, J. C. 2003, A\&A,
408, 353
\bibitem{}Koribalski, B., Johnston, S., Weisberg, J.M., \& Wilson, W. 1995,
  MNRAS, 441, 752
\bibitem{}van der Hucht, K.A. 2001, New Astron. Rev., 45, 135
\bibitem{}Leitherer, C., Chapman, J.M., \& Koribalski, B. 1997, ApJ, 481, 898
\bibitem{}Manchester, R.N., Robinson, B.J., \& Goss, W.M. 1970, Aust. J. Phys., 23, 751
\bibitem{}McClure-Griffiths, N.M., Green, A.J., Dickey, J.M., et al. 2001,
  ApJ, 551, 394
\bibitem{}Mereghetti, S., Belloni, T., Shara, M. \& Drissen, L. 1994,
  ApJ, 424, 943
\bibitem{}Moffat, A.F.J., Shara, M.M., \& Potter, M. 1991, AJ, 102, 642
\bibitem{}Morras, R., \& Cappa, C.E. 1995, IAR Internal Report No. 74
\bibitem{}M\"ucke, A. \& Pohl, M. 2002, in ``Interacting Winds from Massive
  Stars'', eds. A.F.J. Moffat, \& N. St-Louis, ASP Conf. Ser. 260, p. 355
\bibitem{}Niemela, V., et al. 2005, in preparation
\bibitem{}Nolan, P.L., Tompkins, W.F., Grenier, I.A., \& Michelson, P.F. 2003,
  ApJ, 597, 615
\bibitem{}Nugis, T., \& Lamers, H.J.G.L.M. 2000, A\&A, 360, 227
\bibitem{}Reig, P. 1999, A\&A, 345, 576
\bibitem{}Pollock, A.M.T. 1987, A\&A, 171, 135
\bibitem{}Pollock, A.M.T., Corcoran, M.F., Stevens, I.R., \& Williams, P.M.
  2005, ApJ, in press
\bibitem{}Prinja, R.K., Barlow, M.J., \& Howarth, I.D. 1990, ApJ, 361, 607
\bibitem{}Roberts, M.S.E., Romani, R.W., \& Kawai, N. 2001, ApJS,
133, 451
\bibitem{}Rodgers, A.W., Campbell, C.T., \& Whiteoak, J.B. 1960, MNRAS, 121, 103
\bibitem{}Romero, G.E. 2001, in ``The Nature of Unidentified Galactic
  High-Energy Gamma-Ray Sources'', eds. A. Carrami\~nana, O. Reimer, \& D.J.
  Thompson, Kluwer, 267, p. 67
\bibitem{}Romero, G.E., Benaglia, P., \& Torres, D.F. 1999, A\&A, 348, 868
\bibitem{}Skinner, S. L., Zhekov, S. A., G\"udel, M., \& Schmutz, W.
2002a, ApJ, 572, 477
\bibitem{}Skinner, S. L., Zhekov, S. A., G\"udel, M., \& Schmutz, W.
2002b, ApJ, 579, 764
\bibitem{}Stevens, I.R., Blondin, J.M., \& Pollock, A.M.T. 1992, ApJ, 386, 265
\bibitem{}Torres, D.F., Romero, G.E.,  Combi, J.A., \& Benaglia, P. 2001,
  A\&A, 370, 468
\bibitem{}Vacca, W.D., Garmany, C.D., \& Schull, J.M. 1996, ApJ, 460, 914
\bibitem{}Vink, J.S., de Koter, A., \& Lamers, H.J.G.L.M. 2000, A\&A, 362, 295
\bibitem{}V\"olk, H.J., \& Forman, M. 1982, ApJ, 253, 188
\bibitem{}Wackerling, L.R. 1969, Mem. RAS, 73, 153
\bibitem{}White, R.L. 1985, ApJ, 289, 698
\bibitem{}White, R.L., \& Chen, W. 1992, ApJ, 387, L81
\bibitem{}Whiteoak, J.B.Z., \& Uchida, K.I. 1997, A\&A, 317, 563
\bibitem{}Williams, P.M., van der Hucht, K.A., Pollock, A.M.T, et al. 1990,
   MNRAS, 243, 662
\end{thebibliography}
\end{document}